%% file: paper_GN.tex
\documentclass[12pt]{article}

\usepackage[latin1]{inputenc}
\usepackage[english]{babel}
\usepackage{latexsym}
\usepackage{amssymb}
\usepackage{graphicx}
\usepackage{amscd}
\usepackage{amsfonts,epsfig}

\input{TXSsymb}

\newcommand{\di}{\mathrm{d}}

\newcommand{\ol}{\overline}

\newcommand{\vp}{\mathbf{p}}
\newcommand{\vq}{\mathbf{q}}

\newcommand{\LL}{\left}
\newcommand{\RR}{\right}

\begin{document}

\title{Large-$N_f$ chiral transition in the Yukawa model} 

\author{
  {\small Sergio Caracciolo}              \\[-0.2cm]
  {\small\it Dipartimento di Fisica dell'Universit\`a di Milano, 
           I-20133 Milano} \\[-0.2cm]
  {\small\it INFN, Sez. di Milano, ITALIA}    \\[-0.2cm]
  {\small {\tt Sergio.Caracciolo@mi.infn.it}}     \\[-0.2cm]
  \\[-0.35cm] \and
  {\small Bortolo Matteo Mognetti}              \\[-0.2cm]
  {\small\it Dipartimento di Fisica dell'Universit\`a di Milano, 
          I-20133 Milano} \\[-0.2cm]
  {\small\it INFN, Sez. di Milano, ITALIA} \\[-0.2cm]
  {\small {\tt Bortolo.Mognetti@mi.infn.it}}     \\[-0.2cm]
  \\[-0.35cm] \and
 {\small Andrea Pelissetto }        \\[-0.2cm]
  {\small\it Dipartimento di Fisica and INFN -- Sezione di Roma I}  \\[-0.2cm]
  {\small\it Universit\`a degli Studi di Roma ``La Sapienza''}      \\[-0.2cm]
  {\small\it I-00185 Roma, ITALIA}        \\[-0.2cm]
  {\small {\tt Andrea.Pelissetto@roma1.infn.it}}  \\[-0.2cm]
  {\protect\makebox[5in]{\quad}}  
   \\
}

\maketitle


\begin{abstract}
We investigate the finite-temperature behavior of the 
Yukawa model in which $N_{f}$ fermions 
are coupled with a scalar field $\phi$ in the limit $N_f \to \infty$.
Close to the chiral transition the model shows a crossover between 
mean-field behavior (observed for $N_f = \infty$) and Ising behavior 
(observed for any finite $N_f$). We show that this crossover is universal
and related to that observed in the weakly-coupled $\phi^4$ theory. 
It corresponds to the renormalization-group flow from the unstable 
Gaussian fixed point to the stable Ising fixed point. This equivalence
allows us to use results obtained in field theory and in medium-range spin
models to compute Yukawa correlation functions in the crossover regime.

\end{abstract}

\clearpage

\section{Introduction}

The finite-temperature transition in QCD has been extensively studied in the
last twenty years and is becoming increasingly important because of the 
recent experimental progress in the physics of ultrarelativistic
heavy-ion collisions. Some general features of the transition, which is 
associated with the restoration of chiral symmetry, can be 
studied in dimensionally-reduced three-dimensional models 
\cite{PW-84,BPV-03}. However, a detailed understanding 
requires a direct analysis in QCD. Being the phenomenon
intrinsically nonperturbative, our present knowledge comes mainly from 
numerical simulations \cite{Karsch-02,KL-03}. Due to the many technical
difficulties---finite-size effects, proper inclusion of fermions, etc.---results
are not yet conclusive and thus it is worthwhile to study  simplified
models that show the same basic features but are significantly 
simpler. In this paper we shall consider a Yukawa model in which
$N_f$ fermions are coupled with a scalar field through a Yukawa 
interaction. The action of the model in $d+1$ dimensions is
\begin{eqnarray}
{\cal S} &=& D N_{f}
   \int\di^{d+1} \mathbf{x}\,\Big({1\over 2}
      (\partial \phi)^2 + {\mu \over 2} \phi^2 
       +{\lambda  \over 4!} \phi^4\Big) 
+\sum_{f=1}^{ N_f} \int\di^{d+1}\mathbf{x}\,
        \ol \psi_{f} \Big(\sc{\partial}+ g \phi+M\Big)\psi_{f},
\label{action}
\end{eqnarray}
where $\tr \gamma_\mu^2 =  D$ ($D = 2^{d/2}$ if $d$ is even, 
$D = 2^{(d+1)/2}$ if $d$ is odd), the integration is over 
$\mathbb{R}^{d}\times [0,T^{-1}]$, and $\lambda \ge 0$ to ensure the 
stability of the quartic potential.  Along the
\emph{thermal} direction we take periodic boundary conditions for the
bosonic field $\phi$ and antiperiodic ones for the fermionic fields
$\psi_f$.
The theory must be properly regularized. We shall consider a 
sharp-cutoff regularization, restricting the momentum integrations in the 
\emph{spatial} directions to $p < \Lambda$. 
However, the discussion presented here can be extended without 
difficulty to any other 
regularization that mantains at least a remnant of chiral symmetry.

In the limit $N_f\to \infty$ this model can be solved analytically 
and one finds that there is a range of parameters in which 
it shows a transition analogous to that observed in QCD \cite{RSY-94,MZ-03}. 
It separates a low-temperature phase in which chiral symmetry
is broken from a high-temperature phase in which chiral symmetry is 
restored. For $N_f=\infty$ this transition shows mean-field behavior, 
in contrast with general arguments that predict the transition to 
belong to the Ising universality class. This apparent contradiction
was explained in Ref.~\cite{KSS-98} where, by means of scaling arguments,
it was shown that the width of the Ising critical region scales as 
a power of $1/N_f$, so that only mean-field behavior can be observed 
in the limit $N_f = \infty$. An analogous behavior was 
observed in a generalized $O(N)$ $\sigma$ model in
Ref.~\cite{CP-02}: for finite values of $N$ the transition was expected 
to be in the Ising universality class, while the $N=\infty$ solution
predicted mean-field behavior. In Ref.~\cite{CMP-05} we performed a 
detailed calculation of the $1/N$ corrections, explaining the observed 
behavior in terms of a critical-region suppression.
The analytic technique discussed in Ref.~\cite{CMP-05} can be applied to 
model (\ref{action}). It allows us to obtain an analytic description of the 
crossover from mean-field to Ising behavior that occurs when $N_f$ is large
and to extend the discussion of Ref.~\cite{KSS-98} to the case $M\not = 0$.
More importantly, we are able to show that the phenomenon is universal.
In field-theoretical terms, it can be characterized as a crossover 
between two fixed points: the Gaussian fixed point and the Ising fixed point.
This implies that quantitative predictions for model (\ref{action}) 
can be obtained in completely different settings. One can use field theory
and compute the crossover functions by resumming the perturbative series
\cite{BB-85,BBMN-87,BB-02,PRV-99}. Alternatively, one can use the fact that 
the field-theoretical crossover is equivalent to the critical 
crossover that occurs in models with medium-range interactions
\cite{LBB-96,PRV-98,PRV-99,CCPRV-01,PV-02}. This allows one to use the 
wealth of results available for these spin systems
\cite{LBB-96,LBB-97,LB-98,PRV-98,PRV-99,PV-02}. In this case the interaction
range $R$ is essentially equivalent to a power of ${N}$, $N\sim R^d$.
Finally, we should note that the phenomenon is quite general and occurs in any 
situation in which there is a crossover from the Gaussian fixed point to a 
nonclassical stable fixed point. For instance, similar considerations have been
recently presented for finite-temperature QCD in some very specific limit
\cite{Bringoltz-05}.

The paper is organized as follows. In Sec.~\ref{sec2} we review the 
behavior in the limit $N_f = \infty$. In Sec.~\ref{sec3} we consider the 
$1/N_f$ fluctuations and determine the effective theory of the 
excitations that are responsible for 
the Ising behavior at the critical point. These modes are 
described by an effective weakly-coupled $\phi^4$ Hamiltonian.
In Sec.~\ref{sec4.1} we present a general discussion of 
the critical crossover limit.
These considerations are applied to the Yukawa model in Sec.~\ref{sec4.2}
and in Sec.~\ref{sec4.3}. We determine the relevant scaling variables 
and show how to compute the crossover behavior of the correlation functions.
Finally, in Sec.~\ref{sec5} we present our conclusions. In the appendix 
we discuss the relations among medium-ranged spin models, field theory,
and the Yukawa model. A short summary of this work was presented in 
Ref.~\cite{CMP-05b}.

\section{Behavior for $N_f = \infty$} \label{sec2}

The solution of the model for $N_f = \infty$ is quite standard.
We briefly summarize here the main steps, following the presentation 
of Ref.~\cite{MZ-03}. As a first step we integrate the fermionic fields 
obtaining an effective action ${\cal S}_{\rm eff}[\phi]$ given by
\begin{eqnarray}
e^{-D N_{f}{\cal S}_{\rm eff} [\phi]} 
&=& \int \prod_{f=1}^{N_{f}}\di\ol\psi_{f}\di\psi_{f}\,
e^{-{\cal S}[\phi,\ol\psi,\psi]}, 
\end{eqnarray}
where
\begin{eqnarray}
{\cal S}_{\rm eff}[\phi] &=& \int\di^{d+1} \mathbf{x}
\Big({1\over 2}(\partial\phi) ^2  + {\mu\over 2}  \phi ^2 
      +{ \lambda  \over 4!} \phi^4 \Big)
- {1\over D} \int\di^{d+1} \mathbf{x}\,
  \tr \log \Big(\sc{\partial} + g \phi+M\Big) \; .
\end{eqnarray}
For $N_f\to \infty$ one can expand around the saddle point $\phi = \ol \phi$,
that is determined by the gap equation
\begin{equation}\label{saddle_a}
\ol \mu m+{\ol \lambda \over 6 }m^3 =  (m+M) T \sum_{n\in\mathbb{ Z}}
\int_{p < \Lambda} 
   {\di^{d}\vp \over (2 \pi)^{d}} 
\,{1\over {p}^2 + \omega_n^2 + (m+M)^2},
\end{equation}
where we define the {\em frequencies} $\omega_n \equiv (2 n + 1) \pi T$, and
\begin{eqnarray}
m \equiv  g\,\ol\phi\qquad
\ol\mu \equiv  \mu\,g^{-2}\qquad
\ol\lambda \equiv  \lambda\, g^{-4}\; .
\nonumber
\end{eqnarray}
The action corresponding to a saddle-point solution $m$ is:
\begin{eqnarray}
\overline{\cal S}_{\rm eff} (m,M,T) &=& 
  {\ol\mu\over 2}m^2+{\ol \lambda\over 4!}m^4
-{T\over 2}\sum_{n\in \mathbb{Z}}\int_{p<\Lambda} 
{\di^{d} \mathbf{p}\over (2\pi)^d} 
   \log \left[{{p}^2 + \omega_n^2 +(m+M)^2 \over 
              {p}^2 + \omega_n^2} \right] ,
\label{eq5}
\end{eqnarray}
where we have added a mass-independent counterterm to regularize
the sum \cite{MZ-03}. Such a quantity has been chosen so that 
the action for $M=m=0$ vanishes.
Summations can be done analytically \cite{MZ-03}. 
The gap equation can then be written as 
\begin{eqnarray}\label{gap}
p(m) &=& (m+M)\, {\cal G}(m+M,T),
\end{eqnarray}
where the functions $p(m)$ and
${\cal G}(x,T)$ are defined by
\begin{eqnarray}
p(m) &\equiv &\ol\mu \,m+{\ol \lambda \over 6 }m^3 ,
\label{p} \\
{\cal G}(x,T) &\equiv & \int_{p < \Lambda}
{\di^{d}\vp\over (2\pi)^d} \,
 {1\over \sqrt{p^2 + x^2} }\LL({1\over 2}-{1\over e^{\sqrt{p^2 + x^2}/T}+1} 
  \RR) \; .
\label{defG}
\end{eqnarray}
Analogously we can rewrite Eq.~(\ref{eq5}) as
\begin{eqnarray}
\overline{\cal S}_{\rm eff} (m,M,T) &=& 
    {\ol \mu\over 2 } m^2+{\ol \lambda\over 4! } m^4
\nonumber \\
&& -
  T \int_{p<\Lambda} {\di^{d} \mathbf{p}\over (2 \pi)^d}\,
    \left[
   \log \left(  \cosh {\sqrt{p^2 + (m + M)^2}\over 2T} \right) - 
   \log  \cosh {p \over 2T } \right]\; .
\label{ffe}
\end{eqnarray}
Using Eqs.~(\ref{gap}) and (\ref{ffe}) 
we can determine the phase diagram of the model.
Given $\ol \mu$ and $\ol \lambda$, for each value of $T$ and $M$ we 
determine the solutions $m$ of the gap equations. In general,
we find either one solution or three different solutions 
$m_0$, $m_+$, and $m_-$ with $m_-\le m_0\le m_+$ (for some specific values 
of the parameters two of them may coincide). When the solutions are more 
than one, the physical solution is the one with the lowest 
action $\overline{\cal S}_{\rm eff} (m,M,T)$.

Note that Eqs.~(\ref{gap}) and (\ref{ffe}) are invariant 
under the transformations $m\to -m$ and $M\to -M$. Thus, we can limit our
study to the case $M\ge 0$. There are four different possibilities:
\begin{itemize}
\item[(a)]
If $\ol \mu \ge {\cal G}(0,0) = C_0 \Lambda ^{d-1}$ with
\nonumber\\
\begin{eqnarray}
C_0 &\equiv & \Big[ 2^{d}\pi^{d/2}(d-1)\Gamma\Big({d\over 2} \Big)
\Big]^{-1},
\label{der}
\end{eqnarray}
then, for every $M\ge 0$, there is only one solution $m_+\ge 0$; for 
$M = 0$ we have $m_+=0$.
\item[(b)] 
If $0 < \ol \mu < C_0 \Lambda ^{d-1}$, there exists a critical temperature
$T_c(\ol \mu)$. For $T > T_c(\ol \mu)$ and any $M$ 
there is only one solution $m_+\ge 0$ (for $M = 0$ we have $m_+=0$).
For $T < T_c(\ol \mu)$ and $M < \tilde{M}$  there are three solutions 
$m_0$, $m_+$, and $m_-$ with $m_-\le m_0\le m_+$ and
$m_+\ge0$ and $m_-\le 0$. For $T < T_c(\ol \mu)$ and $M > \tilde{M}$ 
there is only one solution corresponding to $m_+$. 
The physical solution is always $m_+$ so that $\tilde{M}$ has no physical
meaning.
Moreover, for $T < T_c(\ol \mu)$ and $M = 0$, $m_+ > 0$. 
The critical temperature can be computed from the following relation:
\begin{eqnarray}
\ol \mu  = {\cal G}(0,T_c) &=&
  T_c \sum_{n\in\mathbb{ Z}}
\int_{p < \Lambda} 
   {\di^{d}\vp \over (2 \pi)^{d}} 
\,{1\over {p}^2 + \omega_{c,n}^2}
\nonumber \\ 
 &=&  \Lambda^{d-1}C_0 - 
  \Lambda^{d-1}\int_{p< 1}{\di^{d}\vp\over (2\pi)^d}\, 
{1\over p [e^{p/t_c}+1]},
\label{eq:Tc}
\end{eqnarray}
where $\omega_{c,n} \equiv (2 n + 1) \pi T_c$ and 
$T_c(\ol\mu)=t_c(\ol\mu)\Lambda$.
For $\ol\mu\to 0$, we have $T_c(\ol\mu)\to\infty$.
\item[(c)] If $ -C_1 < \ol{\mu} \le 0$, with\footnote{$C_1$ is the solution 
of the equation $p(-x) = \lim_{M\to\infty} M {\cal G}(M,0)$ with
$x = (2 C_1/\overline{\lambda})^{1/2}$. The value $x$ corresponds to the 
position of a maximum of $p(m)$ for $\ol \mu = - C_1$.}
\begin{eqnarray}
C_1^{3/2} \equiv \lambda^{1/2}
{3 \Lambda^{d}\over 2^{d+2} \pi^{d/2} \sqrt{2} } \Gamma\Big({d+2\over 2}
\Big)^{-1},
\label{dis_mu}
\end{eqnarray}   
there is a critical mass $\tilde M$ such that there are
three solutions for  $M< \tilde M$, two of them coincide for $M = \tilde M$,
while for  $M> \tilde M$ the only solution
is $m_+$. The physical solution---the one with the lowest action---is 
always $m_+$ so that
$\tilde M $ has no physical meaning. Note that $m_+> 0$ for $M = 0$.
\item[(d)]  For  $ \ol{\mu} \le - C_1$ there are three solutions 
for all values of $T$ and $M$. The relevant solution is always $m_+> 0$.
\end{itemize}
In case (a) chiral symmetry is never broken, while in cases (c) and (d) 
chiral symmetry is never restored. Thus, the only case of interest---and 
the only one we shall consider in the following---is case (b), in which
there is a chirally-symmetric high-temperature phase and 
a low-temperature phase in which chiral symmetry is broken. 

The nature of the transition is easily determined.
We expand 
\begin{eqnarray}
{\cal G}(x,T) &=& \sum_{m,n} g_{mn}\, x^{2m} (T-T_c)^n,
\label{Gexp}
\end{eqnarray} 
where
\begin{eqnarray}
&& g_{00} = 
  T_c \sum_{n\in\mathbb{ Z}}
\int_{p < \Lambda} 
   {\di^{d}\vp \over (2 \pi)^{d}} 
\,{1\over {p}^2 + \omega_{c,n}^2},
\nonumber \\
&& g_{01} = 
  \sum_{n\in\mathbb{ Z}}
\int_{p < \Lambda} 
   {\di^{d}\vp \over (2 \pi)^{d}} 
\,{{p}^2 - \omega_{c,n}^2 \over ({p}^2 + \omega_{c,n}^2)^2}
\nonumber \\
&& \qquad = - {1\over T_c^2}
  \int_{p < \Lambda} {\di^{d}\vp \over (2 \pi)^{d}} 
  {e^{p/T_c}\over (e^{p/T_c} + 1)^2},
\nonumber \\
&& g_{10} = - T_c \sum_{n\in\mathbb{ Z}}
\int_{p < \Lambda} 
   {\di^{d}\vp \over (2 \pi)^{d}} 
\,{1\over ({p}^2 + \omega_{c,n}^2)^2},
\label{defgab}
\end{eqnarray}
and $\omega_{c,n} \equiv  (2 n + 1)\pi T_c$.
Since $g_{00}= {\cal G}(0,T_c) = \ol \mu$ [Eq.~(\ref{eq:Tc})], 
the gap equation (\ref{gap}) becomes:
\begin{eqnarray}
{\ol \lambda \over 6 } m^3 &=& \ol\mu M + g_{01}(T-T_c)(m+M)+ g_{10}(m+M)^3
+\cdots 
\end{eqnarray}
where we have neglected subleading terms in $m+M$ and $T-T_c$. 
Defining
\begin{eqnarray}
u_h \equiv  {6 \ol \mu \over 6 g_{10} - \ol \lambda}\,  M
&\qquad & 
u_t\equiv {6 g_{01}\over 6 g_{10} - \ol \lambda}\, (T-T_c) ,
\end{eqnarray}
and taking the limit $u_h, u_t \to 0$ at fixed $x\equiv u_t/u_h^{2/3}$
we obtain the equation of state
\begin{eqnarray}
m &=& {u_h}^{1/3} f(x)
\\
0 &=& f(x)^3+x f(x) +1.
\label{eqst}
\end{eqnarray}
Note that the prefactor of $T - T_c$ in $u_t$ is always positive to ensure the 
existence of only one solution for $T > T_c$. 
Such an equation is exactly the mean-field equation of state that 
relates magnetization $\varphi$, magnetic field $h$, and 
reduced temperature $t$.
Indeed, if we consider the mean-field Hamiltonian
\begin{equation}
{\cal H} = h \varphi + {t\over2} \varphi^2 + {u\over 24} \varphi^4,
\end{equation}
the stationarity condition gives
\begin{equation}
h + t \varphi + {u\over6} \varphi^3 = 0,
\end{equation}
which is solved by $\varphi = A h^{1/3} {f}(B t |h|^{-2/3})$, where
${f}(x)$ satisfies Eq.~(\ref{eqst}), and $A$ and $B$ are constants
depending on $u$.
This identification also shows that $M$ plays the role of an external
field, while $m\sim \overline{\phi}$ is the magnetization.

\section{Effective theory for the zero mode}\label{sec3}

In order to perform the $1/N_f$ calculation, we expand the field
$\phi$ around the saddle-point solution, 
\begin{equation}
\phi(\mathbf{x}) = \ol \phi + {1 \over g \sqrt{N}}  \hat{\phi}(\mathbf{x}),
\end{equation}
where $N \equiv  D N_f$, 
and $\hat{\phi}(\mathbf{x})$ in Fourier modes:
\begin{eqnarray}
\hat{\phi}(\mathbf{x}_d, x_{d+1}) = T
\sum_{n\in \mathbb{Z}} e^{2 i\pi n T x_{d+1} }
\int {\di^{d} \vp\over (2\pi)^d} \, \hat{\phi}_n(\vp) 
e^{i\vp\cdot\mathbf{x}_d} \; .
\end{eqnarray}
In the following we will refer to the integers $n$---or more precisely to 
$2\pi n T$---as frequencies.
In this way we obtain the following expansion for the effective action:
\begin{eqnarray}
\hat{\cal S}_{\mathrm{eff}}[\hat{\phi}_n] &=& 
N \{ {\cal S}_{\mathrm{eff}}[\phi] - {\cal S}_{\mathrm{eff}}[\ol\phi]\}
\nonumber \\
&=& {T\over 2}\sum_n\int_{p < \Lambda} 
  {\di^{d} \vp \over (2\pi)^d}
\,P(\vp,n) \hat{\phi}_n(\vp) \hat{\phi}_{-n}(-\vp)
\nonumber\\
&&+\sum_{l \ge 3}{T^{l-1}\over l! N^{l/2-1} }
   \sum_{\{n_i\}}
 \int_{p_i < \Lambda} 
   {\di^{d} \vp_1\over (2\pi)^d} \cdots {\di^{d} \vp_l\over (2\pi)^d}
   (2\pi)^d \delta \Big(\sum_{j=1}^l \vp_{j} \Big)\delta_{\sum_i n_i , 0}
\nonumber\\
&&\qquad \qquad
 \hat{\phi}_{n_1}(\vp_1) \cdots \hat{\phi}_{n_l}(\vp_l)
 V^{(l)}(\vp_{1},n_1;\cdots;\vp_{l},n_l).
\label{action-exp}
\end{eqnarray}
Note the Kronecker $\delta$ on the frequencies that ensures that 
vertices are nonvanishing only for $\sum_i^l n_i = 0$. In particular---this
property will be important below---if only one frequency is nonvanishing,
we have 
$V^{(l)}(\vp_{1},n;\vp_{2},0;\cdots;\vp_{l},0) = 0$.

The vertices appearing in expansion (\ref{action-exp}) 
are easily computed. The fermion contribution is obtained by 
considering the one-loop fermionic graphs in theory (\ref{action}). 
If we define the free fermion propagator
\begin{equation}
\Delta_F(\mathbf{p},n;m) \equiv  
{-i\sum_{j=1}^{d}\gamma_j {p}_j-i
\gamma_{d+1} \omega_n +m \over {p}^2+ \omega_n^2 +m^2}
\end{equation}
with $\omega_n \equiv  (2 n + 1) \pi T$, 
the fermion contribution is 
\begin{eqnarray}
&& V^{(l)}_f(\vp_{1},n_1;\cdots;\vp_{l},n_l) = 
\nonumber \\ 
&& \qquad\qquad
  (-1)^l {T\over D} \sum_{a\in \mathbb{Z}}
    \int_{q < \Lambda} {\di^{d} \vq\over (2\pi)^d}\, \hbox{tr}\, 
   \left[ \prod_{i=1}^l \Delta_F \Bigl(\vq + \sum_{j=1}^i \vp_j;
                      a + \sum_{j=1}^i n_j; m + M\Bigr) \right] 
\nonumber \\
   &&  \qquad\qquad
   \vphantom{{\di^{d} \vq\over (2\pi)^d}}
    + \hbox{permutations},
\end{eqnarray}
where the permutations make the vertex completely symmetric 
[there are $(l-1)!$ terms]. Note that the frequencies $\omega_n$ 
never vanish and thus the vertices have a regular expansion 
in powers of $m + M$. Vertices $V_f^{(l)}$ satisfy an important 
symmetry relation. First, note that 
\begin{equation}
\gamma_{d+1} \Delta_F(\vp,n;m) \gamma_{d+1} = 
- \Delta_F(\vp,-n-1;-m).
\end{equation}
It follows
\begin{eqnarray}
&& \sum_{a\in \mathbb{Z}}\hbox{tr}\,
   \left[\prod_{i=1}^l \Delta_F(\vq_i ; a + b_i; m)\right] = 
\sum_{a\in \mathbb{Z}}\hbox{tr}\,
\left[\prod_{i=1}^l 
   \gamma_{d+1} \Delta_F(\vq_i ; a + b_i; m)\gamma_{d+1}\right]=
\\
&& \qquad 
 (-1)^l \sum_{a\in \mathbb{Z}}\hbox{tr}\,
   \left[\prod_{i=1}^l \Delta_F(\vq_i ; - a - b_i - 1; -m)\right] = 
 (-1)^l \sum_{a\in \mathbb{Z}}\hbox{tr}\,
   \left[\prod_{i=1}^l \Delta_F(\vq_i ; a - b_i; -m)\right].
\nonumber 
\end{eqnarray}
In the second step we used $\gamma_{d+1}^2 = 1$, while in the 
last one we redefined $a \to - a + 1$. This relation implies
(we write here explicitly the dependence of the vertices on $m$ and 
$M$)\footnote{If $d$ is odd, one can repeat the same argument 
using $\gamma_{d+2} = \prod_{i=1}^{d+1}\gamma_i$. It shows that 
vertices with $l$ legs are multiplied by $(-1)^l$ if 
one changes the sign of $m + M$ at fixed momenta and {\em frequencies}.}
\begin{equation}
V_f^{(l)}(\vp_1,n_1;\ldots;\vp_l,n_l;m + M) = (-1)^l
V_f^{(l)}(\vp_1,-n_1;\ldots;\vp_l,-n_l;-m - M).
\label{symmetryVf}
\end{equation}
For every $l > 4$, the vertex is due only to the fermion loops, so that 
$V^{(l)} = V^{(l)}_f$. For $l \le  4$ we must also take into account the 
contribution of the bosonic part of the action, so that 
\begin{eqnarray}
 V^{(3)}(\mathbf{p},n_1;\mathbf{q},n_2;\mathbf{r},n_3) 
&=&  \ol\lambda m + 
 V^{(3)}_f(\mathbf{p},n_1;\mathbf{q},n_2;\mathbf{r},n_3),
\nonumber \\
 V^{(4)}(\mathbf{p},n_1;\mathbf{q},n_2;\mathbf{r},n_3;\mathbf{s},n_4) 
&=&  \ol\lambda + 
 V^{(4)}_f(\mathbf{p},n_1;\mathbf{q},n_2;\mathbf{r},n_3;\mathbf{s},n_4).
\end{eqnarray} 
Finally, for the inverse propagator we have 
\begin{eqnarray}
P(\mathbf{p},n) &=& {\mathbf{p}^2\over g^2}+{(2\pi T n)^2
\over g^2}
 +\ol\mu + {\ol\lambda\over 2} m^2 + 
  V_f^{(2)} (-\vp,-n;\vp,n).
\label{prop_gen}
\end{eqnarray}
Vertices $V^{(l)}$ also satisfy the symmetry relation (\ref{symmetryVf}),
while the inverse propagator $P(\vp,n)$ satisfies 
$P(\vp,n;m,M) = P(\vp,-n;-m,-M)$.
Finally, note that 
$V^{(4)}(\mathbf{0},0;\mathbf{0},0;\mathbf{0},0;\mathbf{0},0)$ is positive 
at the transition. Indeed, one obtains explicitly 
(note that $\lambda \ge 0$ to ensure the stability of the quartic potential)
\begin{equation}
V^{(4)}(\mathbf{0},0;\mathbf{0},0;\mathbf{0},0;\mathbf{0},0) = 
   \overline{\lambda} + 6 T_c \sum_a \int_{q<\Lambda} 
   {\di^{d}\vq \over (2 \pi)^{d}} 
\,{1\over ({q}^2 + \omega_{c,n}^2)^2} > 0.
\label{V4pos}
\end{equation}
It is easy to verify that $P(\mathbf{0},0)$ vanishes at the 
transition. Indeed, for $m = M = 0$ we have
\begin{eqnarray}
P(\mathbf{0},0) &=&  \ol\mu
 -   T\sum_{n\in\mathbb{Z}}\int_{q<\Lambda} {\di^{d} \mathbf{q}\over (2\pi)^d} 
     {1\over {q}^2 + \omega_n^2}
\nonumber\\
&=&  -g_{01} (T-T_c)+O(T-T_c)^2,
\end{eqnarray}
where we used Eqs.~(\ref{Gexp}) and (\ref{eq:Tc}). Thus, the 
mode with $n = 0$ is singular. It is exactly this singularity that forbids 
a standard $1/N_f$ expansion at $T = T_c$ and gives rise
to the Ising behavior. This type of singular behavior is completely
analogous to that observed in Ref.~\cite{CMP-05}. The strategy proposed 
there consists in integrating all the nonsingular modes $\hat{\phi}_n$,
$n\not = 0$, and study the effective theory for the zero mode 
$\hat{\phi}_0$. 

Integrating  all fields $\hat{\phi}_n$ with $n\neq 0$
we obtain the effective action
\begin{equation}
e^{-\tilde{\cal S}_{\mathrm{eff}}[\hat{\phi}_0]}=
\int \prod_{n\neq 0}\di \hat{\phi}_n \,
e^{-\hat{\cal S}_{\mathrm{eff}}[\hat{\phi}_n]},
\end{equation}
with
\begin{eqnarray}
\tilde{\cal S}_{\mathrm{eff}}&=& \sqrt{N} \tilde H\hat{\phi}_0(\mathbf{0})+
    {T\over 2} \int_{q<\Lambda}\,{\di^{d} \mathbf{p}\over (2\pi)^d} \,
\hat{\phi}_0(\mathbf{p})\, \tilde P(\mathbf{p})\,\hat{\phi}_0(-\mathbf{p})
\\
&&+\sum_{l\ge 3}{T^{l-1}\over l! N^{l/2-1}}
\int_{p_i < \Lambda}  
  {\di^{d}\mathbf{p}_1\over (2\pi)^d} \cdots
  {\di^{d}\mathbf{p}_l\over (2\pi)^d}\, 
  (2\pi)^d \delta\Big(\sum_{i=1}^l\vp_i \Big)\, 
\nonumber \\
&& \qquad \times 
  \tilde V^{(l)}(\mathbf{p}_1,\ldots\mathbf{p}_l)
  \hat{\phi}_0(\mathbf{p}_1) \cdots \hat{\phi}_0(\mathbf{p}_l)
\nonumber
\end{eqnarray}
where $H$, $\tilde{P}$, and $\tilde V^{(l)}$ have an expansion in powers of 
$1/N$. The computation of these quantities is quite simple. 
The contribution of order $1/N^k$ to $\tilde V^{(l)}$ is obtained 
by considering all $k$-loop diagrams contributing to 
the $l$-point connected correlation function of $\hat{\phi}_0$ and 
considering only the nonsingular fields (i.e. propagators with 
$n\not=0$) on the internal lines. 
Frequency conservation implies that all tree-level diagrams with more than
one vertex
vanish.\footnote{
For a tree-level graph, the usual topological arguments give the 
relation $\sum_n (n-2) N_n = -2$, where $N_n$ is the number of 
vertices belonging to the graph such that $n$ legs belong to internal
lines. Since $n\ge 1$ if there is more than one vertex, 
the previous equality requires 
$N_1 \ge 2$. But frequency conservation implies that $V_l$ vanishes 
if all frequences but one vanish. Therefore, each nontrivial 
tree-level diagram vanishes.}
Therefore,  $\tilde{H} = O(N^{-1})$, 
$\tilde{P} (\vp) = P(\vp,0) + O(N^{-1})$,
and $\tilde V^{(l)} = V^{(l)}_{n_i=0} + O(N^{-1})$ 
($V^{(l)}_{n_i=0}$ is the vertex $V^{(l)}$ with all frequencies set to zero).
For the inverse propagator $\tilde{P}$ and for the magnetic field $\tilde{H}$
we shall also need the $1/N$ corrections. We obtain
\begin{eqnarray}
\tilde H&=& {H_1\over N} + O(N^{-2})
\\
\tilde P(\mathbf{p}) &=& P(\mathbf{p},0) + {P^{(1)}
(\mathbf{p})\over N} + O(N^{-2}),
\end{eqnarray}
with
\begin{eqnarray}
H_1&=& {T\over 2}\sum_{n\neq 0}\int_{p<\Lambda}
{\di^{d} \mathbf{p}\over (2\pi)^d} \,
V_3(\mathbf{p},n) P(\mathbf{p},n)^{-1} 
\label{H1expr}
\\
P^{(1)}(\mathbf{0})&=& {T\over 2}\sum_{n\neq 0}\int_{p<\Lambda}
{\di^{d} \mathbf{p}\over (2\pi)^d} \,
\,\Big[
 V_4(\mathbf{p},n) P(\mathbf{p},n)^{-1} -  V_{3}(\mathbf{p},n)^2 
 P(\mathbf{p},n)^{-2}
\Big],
\end{eqnarray}
where $V_l(\mathbf{p},n) \equiv  
V^{(l)}(\mathbf{p},n;-\mathbf{p},-n; \mathbf{0},0;\ldots)$. 
Note that relation (\ref{symmetryVf}) implies an analogous 
symmetry relation for $\tilde{V}^{(l)}$:
\begin{equation}
\tilde{V}^{(l)}(\vp_1,\ldots,\vp_l;m,M) = 
(-1)^l \tilde{V}^{(l)}(\vp_1,\ldots,\vp_l;-m,-M),
\label{symmetryVtilde}
\end{equation} 
where we have written explicitly the dependence on $m$ and $M$. 
Analogously $\tilde{P}(\vp)$ and $\tilde{H}$ are respectively 
symmetric and antisymmetric under $m,M\to -m,-M$. 

In order to obtain the final effective theory we introduce a new field
$\chi(\vp)$ such that the corresponding zero-momentum three-leg vertex 
vanishes for any value of the parameters. For this purpose we write
\begin{equation}
\alpha \chi(\vp) = T \hat{\phi}_0(\vp) + \sqrt{N} k \delta(\vp),
\end{equation}
where $\alpha$ and $k$ are functions to be determined. If we write 
$a_l \equiv \tilde{V}^{(l)}(\mathbf{0},0;\ldots;\mathbf{0},0)$, $k$ is 
determined by the equation
\begin{equation}
\sum_{l=0} {(-1)^l k(m,M)^l\over l!} a_{l+3}(m,M) = 0,
\end{equation}
where we have written explicitly the dependence on $m$ and $M$.
Now, symmetry (\ref{symmetryVtilde}) implies also
\begin{equation}
\sum_{l=0} {(-1)^l [-k(-m,-M)]^l\over l!} a_{l+3}(m,M) = 0,
\end{equation}
so that $k(m,M) = - k(-m,-M)$. Therefore, $k$ has an expansion of the 
form
\begin{equation}
k = \sum_{ab,a+b\, {\rm odd}} k_{ab} m^a M^b,
\end{equation}
where the coefficients $k_{ab}$ have a regular expansion in powers of $1/N$.
The leading behavior close to the transition is easily computed:
\begin{equation}
k = {a_3\over a_4} + O(m^a M^b, a + b = 3).
\end{equation}
In terms of $\chi$ the effective action can be written as 
\begin{eqnarray}
\tilde{\cal S}_{\mathrm{eff}}&=& N^{1/2} \ol H\chi(\mathbf{0})+
    {1\over 2} \int_{q<\Lambda}\,{\di^{d} \mathbf{p}\over (2\pi)^d} \,
\chi(\mathbf{p})\, \ol P(\mathbf{p})\,\chi(-\mathbf{p})
\\
&&+\sum_{l\ge 3}{1\over l! N^{l/2-1}}
\int_{p_i < \Lambda}  
  {\di^{d}\mathbf{p}_1\over (2\pi)^d} \cdots
  {\di^{d}\mathbf{p}_l\over (2\pi)^d}\, 
  (2\pi)^d \delta\Big(\sum_{i=1}^l\vp_i \Big)\, 
  \ol V^{(l)}(\mathbf{p}_1,\ldots\mathbf{p}_l)
  \chi(\mathbf{p}_1) \cdots \chi(\mathbf{p}_l).
\nonumber
\end{eqnarray}
The quantities $\ol H$, $\ol P$, and $\overline{V}^{(l)}$ have an expansion
in terms of $m$, $M$, and $1/N$. 
Explicitly we have:
\begin{eqnarray}
&& \ol H = \alpha T^{-1}[ \tilde{H} - k \tilde{P}(\mathbf{0}) + 
             {k^2\over2} \tilde{V}_3(\mathbf{0}) -
             {k^3\over6} \tilde{V}_4(\mathbf{0}) + 
             O(m^a M^b,a+b=5)],
\\
&&\ol P(\vp) = \alpha^2 T^{-1}[ \tilde{P}(\vp) - k \tilde{V}_3(\mathbf{p}) +
             {k^2\over2} \tilde{V}_4(\mathbf{p}) + 
             O(m^a M^b,a+b=4)],
\\
&& {\ol V}^{(2l+1)}(\vp_1,\ldots,\vp_{2l+1}) = 
    \alpha^{2l+1} T^{-1} [ {\tilde V}^{(2l+1)}(\vp_1,\ldots,\vp_{2l+1}) - 
     k {\tilde V}^{(2l+2)}(\vp_1,\ldots,\vp_{2l+1},\mathbf{0}) 
\nonumber \\
&& \qquad\qquad + O(m^a M^b,a+b=3)],
\\
&& {\ol V}^{(2l)}(\vp_1,\ldots,\vp_{2l}) = 
    \alpha^{2l} T^{-1} {\tilde V}^{(2l)}(\vp_1,\ldots,\vp_{2l})  + 
             O(m^a M^b,a+b=2).
\end{eqnarray}
Up to now we have not defined the parameter $\alpha$. We will fix it by
requiring
\begin{equation}
   \left. {d\ol P(\vp)\over dp^2} \right|_{p=0} = 1,
\label{alpha}
\end{equation}
for all values of the parameters. The parameter $\alpha$  is a function 
of $m$, $M$, and $1/N$. The symmetry properties of $k$ and of the vertices
imply that $\alpha$ is invariant under $m,M\to -m,-M$. 
As a consequence, under $m,M\to -m,-M$,
the quantities $\ol H$, $\ol P$, and $\overline{V}^{(l)}$ have 
the same symmetry properties as 
$\tilde H $, $\tilde P$, and $\tilde{V}^{(l)}$. 

In the following we shall need the expansions of 
$\ol H$, $\ol P(\mathbf{0})$, and  
${\ol V}^{(3)}(\mathbf{p},-\mathbf{p},\mathbf{0})$ 
close to the critical point. For this purpose we will use the expansions
\begin{eqnarray}
&& P(\mathbf{0},0)\approx 
   {\ol\lambda \over2} m^2 - (T-T_c) g_{01} - 3 (M + m)^2 g_{10} ,
\nonumber \\
&& V_3(\mathbf{0},0) \approx {\ol\lambda} m - 6 (M+m) g_{10} ,
\nonumber \\[2mm]
&& V_4(\mathbf{0},0) \approx {\ol\lambda} - 6 g_{10},
\label{expan-V2V3V4}
\end{eqnarray}
[$V_n(\vp,0) \equiv  V^{(n)}(\vp,0;-\vp,0;\mathbf{0},0;\mathbf{0},0,\ldots)$]
and the relation
\begin{equation}
V_f^{(3)}(\mathbf{p},n;-\mathbf{p},-n;\mathbf{0},0;m) = 
m V_f^{(4)}(\mathbf{p},n;-\mathbf{p},-n;\mathbf{0},0;\mathbf{0},0;m) + 
   O(m^3),
\label{relV3V4}
\end{equation}
where we have explicitly written the mass dependence of the vertices. 
We expand $\ol H$ and $\ol P(\mathbf{0})$ is powers of $1/N$ as 
\begin{eqnarray}
{T \overline H\over\alpha} = {h}_0 + { {h}_1\over N} + O(N^{-2}), 
\\
{T \overline P(\mathbf{0})\over\alpha^2} = {p}_0 + { p_1\over N} + O(N^{-2}).
\end{eqnarray}
By using expansions (\ref{expan-V2V3V4}) we obtain 
\begin{eqnarray}
{h}_0 &\approx& 
     - {V_3(\mathbf{0},0)\over V_4(\mathbf{0},0)} P(\mathbf{0},0) + 
       {1\over3} {V_3(\mathbf{0},0)^3\over V_4(\mathbf{0},0)^2}
\nonumber \\
    & \approx & - {\ol \mu} M + 
       {g_{01}\ol \lambda\over 6 g_{10} - \ol \lambda} M (T - T_c) - 
       {g_{10}\ol \lambda (6 g_{10} + \ol \lambda)\over 
            (6 g_{10} - \ol \lambda)^2} M^3 + 
        O(m^a M^b,a+b=5),
\nonumber \\
{h}_1 &\approx& 
    H_1 - {V_3(\mathbf{0},0)\over V_4(\mathbf{0},0)} P^{(1)}(\mathbf{0}) 
\nonumber \\
    & \approx & - {\ol \lambda M T \over 2 (6 g_{10} - \ol \lambda)}   
      \sum_{n\neq 0}\int_{p<\Lambda}
      {\di^{d} \mathbf{p}\over (2\pi)^d} \,
      [6 g_{10} + V_f^{(4)}(\mathbf{p},n;-\vp,-n;\mathbf{0},0;\mathbf{0},0)]
      P(\mathbf{p},n)^{-1} 
\nonumber \\
     & &  + O(m^a M^b,a+b=3),
\nonumber \\
{p}_0 &\approx& P(\mathbf{0},0) - 
    {V_3(\mathbf{0},0)^2\over 2 V_4(\mathbf{0},0)} 
\nonumber \\
   &\approx& - g_{01} (T - T_c) + 
      {3 g_{10} \ol \lambda\over 6 g_{10} - \ol \lambda} M^2 + 
       O(m^a M^b,a+b=4),
\nonumber \\
{p}_1 &\approx& P^{(1)} (\mathbf{0},0) \approx e + O(m^a M^b,a+b=2),
\label{expanHP}
\end{eqnarray}
where $e$ is the value of $P^{(1)} (\mathbf{0},0)$ for $M=m=0$.
Note that several terms that are allowed by the symmetry 
$m,M\to -m,-M$ are missing in these expansions. In the case of $h_0$
we used the gap equation to eliminate the term proportional 
to $m^3$. This substitution is responsible for the appearance of the 
term linear in $M$ and cancels the terms proportional 
to $m (T - T_c)$, $m^2 M$, and $m M^2$. In the case of $h_1$ and $p_0$
note that the terms proportional to $m$, and $m^2$, $m M$ cancel out.
Finally, we compute the three-leg vertex. At leading order in $1/N$ 
we obtain 
\begin{eqnarray}
&& {T\over \alpha^3} {\ol V}^{(3)} (\vp,-\vp,\mathbf{0}) = 
            V_3(\vp,0) - {V_3(\mathbf{0},0)\over V_4(\mathbf{0},0)} 
            V_4(\vp,0) 
\nonumber \\
&& \qquad \approx {- \ol \lambda M\over 6 g_{10} - \ol \lambda}
        [6 g_{10} + V_f^{(4)}(\mathbf{p},0;-\vp,0;\mathbf{0},0;\mathbf{0},0)] 
     + O(m^a M^b,a+b=3).
\label{expanV3}
\end{eqnarray}
Note that the term proportional to $m$ is missing as a consequence of 
relation (\ref{relV3V4}).

\section{The critical crossover limit} \label{sec4}

The manipulations presented in the previous section allowed us to 
compute the effective action for the zero mode $\chi(\vp)$. Far from the 
critical point $\ol P(\vp)\not = 0$ for all momenta and thus one can perform
a standard $1/N_f$ expansion. At the critical point instead 
this expansion fails because $\ol P(\mathbf{0}) = 0$. At the critical
point, for $N\to \infty$ the long-distance behavior is controlled by the action
\begin{equation}
\tilde{S}_{\rm eff} \approx 
   \int \di^d {\mathbf x}\, 
   \left[ {1\over2} (\partial \chi)^2 + {u\over 4!} \chi^4 
   \right] + O(N^{-2}),
\label{phi4crit}
\end{equation}
where 
\begin{equation}
u = {1\over N} \overline{V}^{(4)}(\mathbf{0},\mathbf{0},\mathbf{0},\mathbf{0}).
\end{equation}
Here we have used the fact that vertices with an odd number of fields 
vanish at the critical point and the normalization condition (\ref{alpha}).
Moreover, since the critical mode corresponds to $\vp = \mathbf{0}$,
we have performed an expansion in powers of the momenta, keeping only the 
leading term.
Since for $N\to \infty$, 
$\overline{V}^{(4)}(\mathbf{0},\mathbf{0},\mathbf{0},\mathbf{0}) = 
\alpha^4 T^{-1} {V}^{(4)}(\mathbf{0},\mathbf{0},\mathbf{0},\mathbf{0})$,
inequality (\ref{V4pos}) implies $u > 0$ at the critical point. 
Eq.~(\ref{phi4crit}) is the action of the critical $\phi^4$ theory which 
should be studied in the weak-coupling limit $u\to 0$. In this regime
the model shows an interesting scaling behavior---we name it 
{\em critical crossover}---that describes the crossover between mean-field 
and Ising behavior. This allows us to obtain quantitative predictions
for the critical-region suppression observed for $N_f\to \infty$.

\subsection{The general theory} \label{sec4.1}

In this section we wish to review some basic results on the critical 
crossover limit. An extensive discussion can be found in 
Refs.~\cite{BB-85,PRV-99,CMP-05}. Let us first consider the standard
$\phi^4$ theory in $d$ dimensions with $d < 4$.
\begin{equation}
{\cal S}_{\rm cont}[\varphi] = 
\int d^d\mathbf{r}\,
\left[ H \varphi + {{1\over2}} (\partial\varphi)^2 +
    {{r\over2}} \varphi^2 +
    {{u\over 4!}}  \varphi^4\right].
\label{phi4}
\end{equation}
We assume the theory to be regularized with the introduction of a momentum
cutoff. The results are however independent of the chosen regularization 
and one could equally well use a lattice regularization. 
Then, we define a new bosonic field $\psi(\mathbf{s})$ as 
\begin{equation}
\psi(\mathbf{s}) = u^{(d-2)/[2(d-4)]}
    \varphi(\mathbf{s} u^{-1/(4-d)}) .
\end{equation}
Formally, the action can be rewritten as 
\begin{equation}
{\cal S}_{\rm cont}[\psi] =
\int d^d\mathbf{s}\,
\left[ \tilde{H} \psi + {1\over2} (\partial\psi)^2 +
    {\tilde{r}\over2} \psi^2 +
    {1\over 4!}  \psi^4\right],
\end{equation}
where 
\begin{equation}
\tilde{H} \equiv  H u^{-(d+2)/[2(4-d)]}, 
\qquad\qquad
\tilde{r} \equiv  r u^{-2/(4-d)}.
\label{def-tilde}
\end{equation}
Thus, formally, once the action is expressed in terms of 
$\psi$, the bare parameters appear only in the combinations 
$\tilde{H}$ and $\tilde{r}$. 
Then, consider the zero-momentum connected correlation function $\chi_n$. 
We have
\begin{eqnarray}
\chi_n &\equiv &
   \int d^d\mathbf{r}_2\ldots  d^d\mathbf{r}_n\,
   \langle \varphi(\mathbf{0})\varphi(\mathbf{r}_2) \ldots
            \varphi(\mathbf{r}_n) \rangle^{\rm conn}
\nonumber \\
   &=&u^{[2d-n(2+d)]/[2(4-d)]} \int d^d\mathbf{s}_2\ldots  d^d\mathbf{s}_n\,
   \langle \psi(\mathbf{0})\psi(\mathbf{s}_2) \ldots
            \psi(\mathbf{s}_n) \rangle^{\rm conn}
\nonumber \\
      &=& u^{[2d-n(2+d)]/[2(4-d)]} f_n(\tilde{H},\tilde{r}),
\label{scalingrel}
\end{eqnarray}
i.e. $u^{-[2d-n(2+d)]/[2(4-d)]} \chi_n$ is a scaling function of 
$\tilde{H}$ and $\tilde{r}$. Analogously, one can determine the scaling
behavior of the correlation length $\xi$:
\begin{equation}
\xi^2 = {1\over 2d\chi_2} \int d^d\mathbf{r}\, r^2 
             \langle \varphi(\mathbf{0})\varphi(\mathbf{r}) \rangle
      = u^{-2/(4 - d)} f_\xi(\tilde{H},\tilde{r}).
\label{scalingrelxi}
\end{equation}
The above-reported discussion is valid only at the formal level since we have 
not taken into account the presence of the cutoff that breaks scale
invariance. For $d < 4$ only a mass renormalization
(a redefinition of the parameter $r$) is needed in order
to take care of divergencies. By a proper treatment \cite{BB-85} 
one can show that there is a function $r_c(u)$ such that 
the correlation function $\chi_n$ and $\xi$ 
satisfy the scaling relations (\ref{scalingrel}) and 
(\ref{scalingrelxi}) with 
$\tilde{t} = t u^{-2/(4-d)}$, $t \equiv r - r_c(u)$, replacing $\tilde{r}$:
\begin{equation}
  \chi_n = u^{[2d-n(2+d)]/[2(4-d)]} f_n(\tilde{H},\tilde{t}) 
\quad\qquad
  \xi^2 =  u^{-2/(4 - d)} f_\xi(\tilde{H},\tilde{t}).
\label{scalingrel2}
\end{equation}
The function $r_c(u)$ takes care of the ultraviolet divergent diagrams.
In two dimensions, only the tadpole is primitively divergent and we have 
\begin{equation}
r_c(u) = {u\over 8\pi} \ln u + K u.
\end{equation}
In $d = 3$ divergences appear at one and two loops, so that 
\begin{equation}
r_c(u) = - {\Lambda\over 4 \pi^2} u + {u^2\over 96\pi^2} \ln u + K u^2.
\label{rcd3}
\end{equation}
In both cases the arbitrary constant $K$ can be chosen so that 
$t = 0$ corresponds to the critical point. 
The function $r_c(u)$ depends on the chosen regularization
(the expressions we report above correspond to a sharp-cutoff regularization). 
On the other hand, the scaling functions $f(\tilde{H},\tilde{t})$ are 
regularization-independent (universal) once a specific normalization for the fields,
the coupling constant, and the scaling variables is 
chosen.  They are the crossover functions
that relate mean-field and Ising behavior. Consider, for instance, the
case $H = 0$. For $t$ fixed and $u \to 0$ we obtain the standard perturbative
expansion; thus, $\tilde{t}\to \infty$ corresponds to the mean-field limit.
On the other hand, for $t \to 0$ at $u$ fixed, Ising behavior is
obtained; $\tilde{t}=0$ is the nonclassical limit. By varying $\tilde{t}$
between 0 and $\infty$ one obtains the full universal crossover behavior.

In Ref.~\cite{CMP-05} we extended these considerations to the general 
two-dimensional Hamiltonian
\begin{eqnarray}
{\cal S}_{\rm eff}[\varphi] &=& H \varphi(\mathbf{0}) +
      {1\over 2} \int {\di^2 \mathbf{p}\over (2\pi)^2 }\, 
      [K(\mathbf{p}) + r]
      \varphi(\mathbf{p}) \varphi(-\mathbf{p})
\label{ccl} \\
&& +  \sum_{l\ge 3} {u^{l/2-1}\over l!} 
    \int {\di^2 \mathbf{p}_1\over (2\pi)^2 } \ldots 
            {\di^2 \mathbf{p}_l\over (2\pi)^2 }\, 
   (2\pi)^d \delta\left(\sum_i \mathbf{p}_i\right)\, 
                {\cal V}^{(l)}(\mathbf{p}_1,\ldots,\mathbf{p}_l)
                \varphi(\mathbf{p}_1) \ldots
                \varphi(\mathbf{p}_l),
\nonumber 
\end{eqnarray}
where $K(\vp) = p^2 + O(p^4)$, 
${\cal V}^{(3)}(\mathbf{0},\mathbf{0},\mathbf{0}) = 0$, and 
${\cal V}^{(4)}(\mathbf{0},\mathbf{0},\mathbf{0},\mathbf{0}) = 1$. 
The presence of vertices with an odd number of legs requires an 
additional counterterm for the magnetic field. Indeed, we showed that it 
was possible to find functions $r_c(u)$ and $H_c(u)$ such that 
for $t\equiv r - r_c(u)$ (infrared limit), $h\equiv H - H_c(u)$,
$u\to 0$ (weak-coupling limit), at fixed $t/u$, $h/u$ one has 
\begin{equation}
  \chi_n = u^{1-n} f_n(\tilde{h},\tilde{t}),
\end{equation}
where the scaling function $f_n(x,y)$ is the same as that computed in the 
continuum theory. In particular, $\chi_n u^{n-1}$ vanishes in the 
crossover limit if $n$ is odd. The counterterms are regularization-dependent.
In the continuum theory with a cutoff we have 
\begin{eqnarray}
h_c &=& -{\sqrt{u}\over2} \int_{p<\Lambda}
    {\di^{2} \vp \over (2\pi)^2} {{\cal V}_3(\vp)\over K(\vp)},
\\
r_c &=& {u\over 8\pi} \ln {u\over \Lambda^2}  + 
{u\over 2}\int_{p<\Lambda} {\di^{2}\vp \over (2\pi)^2} \,
{{\cal V}_3(\vp)^2\over K(\vp)^2} +  A_0 u,
\label{rc-expr}
\end{eqnarray}
with ${\cal V}_3(\vp) \equiv {\cal V}^{(3)}(\vp,-\vp,\mathbf{0})$ and 
\begin{eqnarray}
A_0&=&-D_2-{3\over 8\pi}+{1\over 8\pi }\log{3 \over 8\pi}-{1\over 2}
\int_{p< \Lambda}{\di^{2}\vp\over (2\pi)^2}\, \left[
   {{\cal V}^{(4)}(\vp,-\vp,\mathbf{0},\mathbf{0})\over K(\vp)}-{1\over p^2}\right].
\nonumber\\
\end{eqnarray}
The nonperturbative constant $D_2$ was estimated in 
Ref.~\cite{PRV-99}: $D_2=-0.0524(2)$.

\subsection{Scaling behavior} \label{sec4.2}

In this section we wish to use the previous results to compute the 
crossover behavior of model (\ref{action}) in 2+1 dimensions.
Since $u\sim 1/N$ the relevant scaling variables are
\begin{eqnarray}
x_h &=& {N T_c\over \alpha} ({N}^{1/2} \ol H - H_c) \nonumber \\
x_t &=& {N T_c\over \alpha^2} (\ol P({\mathbf 0}) - r_c),
\label{scal}
\end{eqnarray}
where the factors $\alpha/T_c$ and $\alpha^2/T_c$ are introduced for convenience. 
The critical crossover limit is obtained by tuning 
$T$, $M$, and $N$ close to the critical point so that 
$x_h$ and $x_t$ are kept constant. 
The expansions of $\ol H$  and $\ol P({\mathbf 0})$ are reported in
Eq.~(\ref{expanHP}). The expansions of $H_c$ and $r_c$ are easily
derived. For $H_c$ we have 
\begin{equation}
H_c = - {1\over 2 \sqrt{N}} 
     \int_{p<\Lambda} {\di^2 \vp\over (2\pi)^2} \,
          {{\ol V}^{(3)}(\vp,-\vp,\mathbf{0}) \over 
      \ol P(\vp)} ,
\end{equation}
where all quantities are computed for $M=m=0$. Using
Eq.~(\ref{expanV3}) we have 
\begin{equation}
H_c = {\alpha h_{c0} M\over T_c \sqrt{N}} + O(m^a M^b,a+b=3),
\end{equation}
where $h_{c0}$ is a constant.
Using Eq.~(\ref{rc-expr}) we obtain for $r_c$ the expansion
\begin{equation}
r_c = {\alpha^2\over T_c N} (r_0 \ln N + r_1) + O(m^a M^b,a+b=2),
\end{equation}
where 
\begin{eqnarray}
r_0 &=& - {\alpha^2 V_4(\mathbf{0},0)\over 8\pi},
\\
r_1 &=& \alpha^2 V_4(\mathbf{0},0) \left[
    {1\over 8\pi} \ln {3 \alpha^4 V_4(\mathbf{0},0) \over 8\pi T_c\Lambda^2}
    - D_2 - {3\over 8\pi}\right]
\nonumber \\
    && 
    - {1\over2} \int_{p<\Lambda} {\di^2 \vp\over (2\pi)^2} \, 
      \left[ {T_c V_4(\vp,0)\over P(\vp,0)} - 
             {\alpha^2 V_4(\mathbf{0},0) \over p^2} \right].
\end{eqnarray}
Note that the three-leg vertex that appears in Eq.~(\ref{rc-expr}) 
does not contribute to this order, since it vanishes for $m=M=0$.
Thus, Eqs.~(\ref{scal}) can be written as 
\begin{eqnarray}
x_h &\approx & N^{3/2} \left[ - \ol \mu M + 
     a_0 M (T - T_c) + a_1 M^3 + {a_2 M\over N} + \cdots \right] - 
      h_{c0} M \sqrt{N},
\\
x_t &\approx & N \left[-g_{01} (T-T_c) + a_3 M^2 + {e\over N} + \cdots
      \right] - 
            r_0 \ln N - r_1,
\end{eqnarray}
where $a_0$, $a_1$, $a_2$, $a_3$, and $e$ are coefficients that can be read from 
Eq.~(\ref{expanHP}). These expansions show that 
\begin{eqnarray} 
&&\ol \mu M = - {x_h N^{-3/2}},
\label{scalM}
\\
&& T-T_c  = {e - r_0 \ln N - r_1 - x_t\over N g_{01}}.
\end{eqnarray}
The critical point is specified by the condition $x_t = x_h = 0$. 
The symmetry under $m,M\to -m,-M$ guarantees that the critical
point corresponds to $M = 0$. On the other hand, 
$1/N$ fluctuations give rise to a shift of the critical temperature.
If $T_c(N)$ is the finite-$N$ critical temperature, we obtain 
\begin{equation}
T_c(N) \approx T_c + {1\over N} {e - r_0 \ln N - r_1\over g_{01}}.
\end{equation}
Note that, beside the expected $1/N$ correction there is also a 
$\ln N/N$ term that is related to the nontrivial renormalization.
It follows
\begin{equation}
T - T_c(N) = - {x_t\over N g_{01}}.
\label{scalT}
\end{equation}
Note that $g_{01}$ is negative [see Eq.~(\ref{defgab})]
and thus we have $x_t > 0$ for $T > T_c(N)$, as expected.
Using the gap equation we can also derive the behavior of $m$ in the 
critical crossover limit.  We obtain 
\begin{eqnarray}
m \equiv {m_0 \over N^{1/2}}, \qquad
\end{eqnarray}
where $m_0$ is a function of $\ln N$ that satisfies the equation
\begin{equation}
{1\over 6} ({\ol\lambda - 6 g_{10}}) m_0^3 + 
(r_0 \ln N  + r_{1} -e + x_t) m_0 + x_h = 0.
\label{eqperm0}
\end{equation}
For $N\to \infty$, $m_0$ has an expansion in inverse powers of $\ln N$,
the leading term being
\begin{equation}
m_0 \approx - {x_h \over r_0} {1\over \ln N} + 
    O(\ln^{-2} N).
\end{equation}
Note that $m_0\to 0$ as $x_h \to 0$.

These results confirm the scaling predictions of Ref.~\cite{KSS-98}. 
For the massless theory with $M = 0$, there are two regimes: 
for $N(T-T_c(N))\ll 1$ one observes Ising behavior, while for 
$N(T-T_c(N))\gg 1$ mean-field behavior occurs. If $M\not=0$ the 
same considerations apply, the relevant variable being 
$M N^{3/2}$. It is important to note the role played in the derivation 
by the symmetry $m,M\to -m,-M$, that is present because the 
regularization preserves chiral invariance. Even though vertices with an
odd number of legs are present, the symmetry makes them irrelevant in the 
crossover limit. Thus, the additional renormalizations computed in 
Ref.~\cite{CMP-05} do not play any role here.

The results reported above can be extended to $d$ dimensions for $d < 4$.
Eq.~(\ref{def-tilde}) implies that the relevant scaling variables are
\begin{equation}
x_h \sim M N^{(d + 2)/[2(4-d)]},\qquad\qquad
x_t \sim [T-T_c(N)] N^{2/(4-d)}.\qquad\qquad
\label{scaling-dgen}
\end{equation}
In $d = 3$, on the basis of Eq.~(\ref{rcd3}),
we also predict for $T_c(N)$ an expansion of the form
\begin{equation}
T_c(N) \approx T_c + {a\over N} + {b\ln N + c\over N^2},
\end{equation}
where $a$, $b$, and $c$ are constants that can be computed as in the 
two-dimensional case.

\subsection{Correlation functions} \label{sec4.3}

The results reported in Sec.~\ref{sec4.2} allow us to compute the 
scaling behavior of the correlation functions. For instance,
we have
\begin{equation}
\langle \phi(\mathbf{x}_d,x_{d+1}) \rangle 
 = {\ol \phi} - {k \over g} + {\alpha\over g \sqrt{N}} 
   \langle \chi(\mathbf{x}_d) \rangle
\end{equation}
Using Eq.~(\ref{scalingrel2}) with $n = 1$ and $d = 2$, we have 
$\langle \chi(\mathbf{x}_d) \rangle = f_1(x_h,x_t)$ in the 
critical crossover limit.
The background term can be neglected in the crossover limit since
\begin{equation}
{\ol \phi} - {k\over g} \approx 
   {1\over g} \left[m - {V_3(\mathbf{0},0)\over V_4(\mathbf{0},0)}\right] 
   \approx 
    {M\over g} {6 g_{10} \over \ol\lambda - 6 g_{10}} \sim N^{-3/2}.
\end{equation}
Thus, we obtain 
\begin{equation}
\langle \phi(\mathbf{x}_d,x_{d+1}) \rangle \approx
    {\alpha\over g \sqrt{N}} f_1(x_h,x_t)
\end{equation}
The factor $1/\sqrt{N}$ is related to the particular normalization 
of $\phi$ used in (\ref{action}) and disappears if we redefine 
$\varphi = \sqrt{N}\phi$ in order to have a canonical kinetic term for 
$\varphi$. The function $f_1(x_h,x_t)$ is the scaling function for the 
magnetization in the Ising model. For instance, for $x_h = 0$ and 
$x_t < 0$ (low-temperature phase), we have 
$f_1(0,x_t) \sim (-x_t)^{\beta_I}$ and 
$f_1(0,x_t) \sim (-x_t)^{\beta_{MF}}$ respectively for 
$|x_t| \ll 1$ and $|x_t| \gg 1$, where $\beta_I = 1/8$ and 
$\beta_{MF} = 1/2$ are the magnetization exponents in the 
Ising and in the Gaussian model. The universality of the crossover 
allows us to compute the scaling functions in any other model in which 
there exists a crossover between the Gaussian and the Ising fixed point.
In particular, we can use the results for systems with medium-range 
interactions \cite{LBB-96,LBB-97} (see also the appendix). 
In Ref.~\cite{LBB-97} (LBB) the authors 
report $\langle |m| R\rangle$ versus $t R^2$ (see their Fig. 9), where 
$m$ is the magnetization, $t$ the reduced temperature, and $R$
the effective interaction range. These results give us
$\langle \phi(\mathbf{x}_d,x_{d+1}) \rangle$ for $x_h = 0$. One only needs
to take into account the different normalizations of the fields, of the 
coupling constant, and of the scaling variable. 
In the crossover limit $N\to\infty$, $T\to T_c(N)$ at
fixed $N(T - T_c(N))$ we have
\begin{eqnarray}
&& g \sqrt{N}\langle \phi(\mathbf{x}_d,x_{d+1}) \rangle  = 
K_{1,\rm LBB} \langle |m| R\rangle_{\rm LBB} \\
&& (t R^2)_{\rm LBB} = K_{\rm LBB} N[T - T_c(N)].
\end{eqnarray}
The nonuniversal constants $K_{\rm LBB}$ and  $K_{1,\rm LBB}$ are 
computed in the Appendix.

It is customary to define an 
effective exponent $\beta_{\rm eff}(T)$ as
\begin{equation}
\beta_{\rm eff}(T) = [T - T_c(N)] {d\over dT} \ln 
   \langle \phi(\mathbf{x}_d,x_{d+1}) \rangle,
\end{equation}
for $M = 0$ and $T < T_c(N)$. In the crossover limit 
$T\to T_c(N)$, $N\to\infty$ at fixed $N[T - T_c(N)]$, 
the exponent $\beta_{\rm eff}(T)$ interpolates 
between the Ising value $\beta_I = 1/8$ and the mean-field $\beta_{MF} = 1/2$.
Again, this effective exponent can be derived from the results of 
Ref.~\cite{LBB-97}.
The curve reported in Fig.~15 of Ref.~\cite{LBB-97}
gives $\beta_{\rm eff}$ in the Yukawa model once $t R^2$ is replaced by 
$K_{\rm LBB} [T - T_c(N)] N$. 

The same considerations apply to the connected zero-momentum $n$-point
function $\chi_n$:
\begin{eqnarray}
\chi_n &=& 
  \int \di^{d+1}\mathbf{x}_2 \ldots  \di^{d+1}\mathbf{x}_n \, 
  \langle \phi(\mathbf{0})\phi(\mathbf{x}_2)\ldots 
          \phi(\mathbf{x}_n) \rangle^{\rm conn} 
\nonumber \\ 
   &=&
  {\alpha^n\over T^{n-1} g^n N^{n/2}} 
  \int \di^{d}\mathbf{x}_2 \ldots  \di^{d}\mathbf{x}_n \, 
  \langle \chi(\mathbf{0})\chi(\mathbf{x}_2)\ldots 
          \chi(\mathbf{x}_n) \rangle^{\rm conn} 
\nonumber \\ 
&=& {\alpha^{4-3n} V_4(\mathbf{0},0)^{1-n} \over g^n} N^{n/2-1} f_n (x_h,x_t),
\end{eqnarray}
For $n = 2$ the crossover function for $x_h = 0$ can be obtained from the 
results of Ref.~\cite{LBB-97}, since 
$g^2 \chi_2 = K_{2,\rm LBB} (\tilde{\chi} R^2)_{\rm LBB}$. The constant
$K_{2,\rm LBB}$ is given in the Appendix.

One can also use field theory to compute the crossover curves and 
thus use the results of Ref.~\cite{PRV-99}.
For instance, in the high-temperature phase, for $M = 0$ we have 
in the crossover limit 
\begin{equation}
g^2 \chi_2 = K_{2,\rm FT} F_\chi (\tilde{t}) ,\qquad\qquad
\tilde{t} = K_{\rm FT} N [T - T_c(N)],
\end{equation}
where $F_{\chi}(\tilde{t})$ is reported in Ref.~\cite{PRV-99} and 
$K_{\rm FT}$, $K_{2,\rm FT}$ are nonuniversal constants computed in the 
appendix.

In the discussion presented above we have focused on the case $d=2$, 
but it is immediate to generalize all these considerations to the 
three-dimensional case. For $d=3$ the universal crossover curves have 
been computed in Refs.~\cite{BB-85,BBMN-87,PRV-99,BB-02} (field theory) and 
in Ref.~\cite{LB-98} (medium-range models). These results apply directly to 
the Yukawa model.

\section{Conclusions} \label{sec5}

In this paper we have considered the Yukawa model in the limit 
$N_f\to \infty$, focusing on the crossover between mean-field and 
Ising behavior. For this purpose we have determined the action of the 
mode that becomes critical at the transition. In the long-distance limit, 
it becomes equivalent to that of a weakly coupled $\phi^4$ theory. This 
identification allows us to use the results available 
for this model \cite{BB-85,CMP-05} and, in particular, to identify 
a universal critical crossover occurring for $N_f\to \infty$, 
$M\to 0$, and $T - T_c(N) \to 0$ at fixed $x_t$ and $x_h$, see
Eqs.~(\ref{scalM}), (\ref{scalT}), (\ref{scaling-dgen}). 
In field-theoretical terms,
this behavior represents the crossover induced by the flow from the 
unstable Gaussian fixed point to the stable Ising fixed point. 
Quantitative results for the Yukawa model can be obtained by using 
the field-theoretical results of Refs.~\cite{BB-85,PRV-99,BB-02},
or the Monte Carlo results available for medium-ranged models
\cite{LBB-97,LB-98}. The necessary nonuniversal renormalization constants 
can be computed in perturbation theory. Results for $d = 2$ are 
reported in the appendix. 

We should stress that our results are not specific of the chosen regularization,
but can be extended to other regularizations as well. In particular,
the extension to Kogut-Susskind fermions, 
the model considered in Ref.~\cite{KSS-98}, is essentially straighforward.
The Wilson case is more involved. Indeed, the absence of chiral symmetry 
implies that the symmetry relations satisfied here by the effective 
vertices [see Eq.~(\ref{symmetryVf})] are no longer valid. In turn, this 
may imply additional mixings as it happens in the generalized
Heisenberg model \cite{CMP-05}.

Let us note that all calculations presented here refer 
to the model in infinite spatial volume. However, the crossover 
behavior can also be observed in the finite-size scaling limit. The discussion
in Sec.~\ref{sec4.1} can be easily extended to this case too. It is trivial
to verify that the correct scaling variable is $\tilde{L} = L u^{1/(4-d)}$, i.e.
$\tilde{L} = L N^{-1/(4-d)}$ in the Yukawa model. Again, one can use universality
and obtain predictions for the Yukawa model from the results obtained in other 
contexts. In particular, one can use the finite-size scaling results of 
Refs.~\cite{LBB-96,LBB-97,LB-98} that refer to medium-range models at the 
critical point.

Finally, we should mention that one could also generalize the model and consider 
fermion fields $\psi_{\alpha f}$ transforming according to a representation of 
a group $G$  and a coupling of the form $\ol\psi_f T^a \phi^a \psi_f$, where 
$T^a$ are the generators of the algebra of $G$. 
The discussion is 
essentially unchanged, though in this case one would obtain the 
vector $\phi^4$ theory. Field-theory results relevant for this case are given in 
Refs.~\cite{BB-85,PRV-99,BB-02}.

\appendix

\section{Relations  among the Yukawa model, medium-range models, 
and field theory} \label{App}

In this appendix we relate the weakly coupled
$\phi^4$ theory, medium-range models, and the Yukawa model for $d=2$.
For simplicity, we only consider the case $H=0$, corresponding to 
$M = 0$ in the Yukawa model. The field-theory model has been
discussed in Sec.~\ref{sec4.1}, where it was shown that 
the $n$-point zero momentum connected correlation function $\chi_{{\rm FT},n}$
shows a scaling behavior of the form
\begin{equation}
u^{n-1} \chi_{{\rm FT},n} = f_{{\rm FT},n}(\tilde{t}_{\rm FT})
\qquad\qquad 
\tilde{t}_{\rm FT} \equiv (r - r_c(u))/u.
\end{equation}
Next, we consider systems with medium-range interactions. Consider a square
lattice, Ising spins $\sigma_x$ at the sites of the lattice, 
and the Hamiltonian
\begin{equation}
{\cal H} = - {1\over2} \sum_{\mathbf{x}\mathbf{y}} 
    J(\mathbf{x}-\mathbf{y}) \sigma_\mathbf{x} \sigma_\mathbf{y}.
\end{equation}
We assume\footnote{One can also consider a much more general class of 
medium-ranged models, see Ref.~\cite{PRV-99}.} that 
$J(\mathbf{x}) = 1$ for $|x| \le  R_m$, $J(\mathbf{x}) = 0$ for $|x| > R_m$. 
The behavior of these
models is very similar to that observed in the Yukawa model, $R_m$ playing the 
role of $N$. For any finite $R_m$, the system belongs to the Ising universality
class, while for $R_m = \infty$ all spins are coupled together and 
one obtains mean-field behavior. In Ref.~\cite{LBB-96} it was shown that 
this model shows a crossover that interpolates between mean-field and Ising
behavior. If one defines an effective interaction range $R$ by 
\begin{equation}
R^2 = {\sum_{\mathbf{x},\mathbf{y}} (\mathbf{x} - \mathbf{y})^2 
        J(\mathbf{x} - \mathbf{y}) \over 
       \sum_{\mathbf{x},\mathbf{y}} 
        J(\mathbf{x} - \mathbf{y}) },
\end{equation}
then for $R,R_m\to \infty$, $t\equiv (T - T_c(R))/T_c(R) \to 0$
at fixed $\tilde{t}_{\rm MR} \equiv  t R^2$ one has 
\begin{equation}
   R^{4-3n} \chi_{{\rm MR},n} = f_{{\rm MR},n}(\tilde{t}_{\rm MR}),
\end{equation}
where $\chi_{{\rm MR},n}$ is the connected zero-momentum $n$-point correlation
function of the fields $\sigma$. In Ref.~\cite{PRV-99} it was shown 
that $f_{{\rm MR},n}(x)$ and $f_{{\rm FT},n}(x)$ are closely related. 
Indeed, we have 
\begin{equation}
    f_{{\rm MR},n}(x) = \mu_{1,\rm MR} \mu_{2,\rm MR}^n 
   f_{{\rm FT},n}(\lambda_{\rm MR} x),
\end{equation}
where $\mu_{i,\rm MR}$ and $\lambda_{\rm MR}$ are model-dependent constants
that reflect the arbitrariness in the definitions of the fields, of the range 
$R$, and of the scaling variable $\tilde{t}$. 
The constants can be computed using the results of Ref.~\cite{PRV-99},
Sec. 4.2.\footnote{Note that the function 
$f_\chi(\hat{t})$ defined in Ref.~\cite{PRV-99} refers to correlations of the 
fields $\phi$ and not of the original fields $\varphi$. However, 
relation (4.12) of Ref.~\cite{PRV-99} 
shows that in the critical crossover limit 
$\sum_x \langle \varphi_0\varphi_x\rangle \approx 
\sum_x \langle \phi_0\phi_x\rangle$. The same holds for $\chi_4$.
The expression reported here are obtained from those reported in 
Ref.~\cite{PRV-99} by setting $\overline{a}_2 = 1$, $\overline{a}_4 = -2$,
$N=1$, $c_0 = \hat{c}_0 = \tau$, and $\tilde{t}_{\rm MR} = \hat{t} + \hat{c}_0$.}
Explicitly, we have for the two-point and four-point zero-momentum connected 
correlation functions:
\begin{eqnarray}
\chi_{{\rm MR},2} R^{-2} &=& {1\over \tilde{t}_{\rm MR}} + 
    {1\over 4 \pi \tilde{t}^2_{\rm MR}} 
    \left[ \ln \left({4\pi \tilde{t}_{\rm MR}\over3}\right) + 8 \pi D_2 + 3 
    \right] + 
    O(\tilde{t}^{-3}_{\rm MR}),
\nonumber \\
\chi_{{\rm MR},4} R^{-8} &=& -{2\over \tilde{t}^4_{\rm MR}} + 
    O(\tilde{t}^{-5}_{\rm MR}).
\end{eqnarray}
In the field-theory model we have instead
\begin{eqnarray}
u \chi_2 &=& {1\over \tilde{t}} + 
     {1\over 8\pi \tilde{t}^2} \left[ \ln {8 \pi \tilde{t}\over 3} + 
           8 \pi D_2 + 3 \right] + O(\tilde{t}^{-3}).
\\
u^2 \chi_4 &=& -{1\over \tilde{t}^4} + O(\tilde{t}^{-5}).
\end{eqnarray}
Comparing we obtain
\begin{equation}
\mu_{1,\rm MR} = 2, \qquad\qquad
\mu_{\rm MR} = \lambda_{\rm MR} = {1\over2}.
\end{equation}
In Sec.~\ref{sec4} we have shown a similar relation for the Yukawa model.
If $x_t \equiv  - g_{01} N [T - T_c(N)]$, we find for the zero-momentum
correlation functions of the field $\phi$ 
\begin{equation}
\tilde{\chi}_n \equiv  g^n \chi_n = 
    N^{n/2-1} f_{Y,n}(x_t),
\end{equation}
and 
\begin{equation}
    f_{Y,n}(x) = \mu_{1,Y} \mu_{2,Y}^n f_{{\rm FT},n}(\lambda_{Y} x).
\end{equation}
In order to compute these constants we compare the one-loop expansions of
the two-point function in field theory and in the Yukawa model.
In the Yukawa model we find
\begin{eqnarray}
\int \di^d \mathbf{x} \langle \chi(\mathbf{0}) \chi(\mathbf{x}) \rangle
   &=& {N T\over \alpha^2 x_t} - \left({N T\over \alpha^2 x_t}\right)^2 r_c 
 \nonumber \\
   && - {1\over 2N} \left({N T\over \alpha^2 x_t}\right)^2
      \int_{p<\Lambda} {\di^2 \vp\over (2\pi)^2} 
      {\overline{V}^{(4)}(\vp,-\vp,\mathbf{0},\mathbf{0}) \over 
       \ol{P}(\vp)} + O(x_t^{-3}).
\end{eqnarray}
Using the explicit expression for $r_c$ we obtain 
\begin{eqnarray}
\tilde{\chi}_2 = {1\over x_t} + 
     {\alpha^2 V_4(\mathbf{0},0) \over 8\pi x_t^2} 
     \left[ \ln {8 \pi x_t\over 3 \alpha^2 V_4(\mathbf{0},0)} + 
            8 \pi D_2 + 3 \right] + O(x_t^{-3}).
\end{eqnarray}
For the four-point function we have instead
\begin{equation}
\tilde{\chi}_4 N^{-1} = - {V_4(\mathbf{0},0)\over x_t^4} 
            + O(x_t^{-5}).
\end{equation}
Comparing we obtain 
\begin{equation}
\mu_{1,Y} = \alpha^4 V_4(\mathbf{0},0), \qquad\qquad
\mu_{2,Y} = {1\over \alpha^3 V_4(\mathbf{0},0)}, \qquad\qquad
\lambda_Y = {1\over \alpha^2 V_4(\mathbf{0},0)}.
\end{equation}
The constants reported in Sec.~\ref{sec4.3} are easily derived:
\begin{eqnarray}
&& K_{\rm FT} = - \lambda_Y g_{01}, \qquad \qquad 
   K_{2,\rm FT} = \mu_{1,Y} \mu_{2,Y}^2 , 
\\
&& K_{\rm LBB} = - {\lambda_Y g_{01}\over \lambda_{\rm MR}}, \qquad \qquad 
   K_{n,\rm LBB} = {\mu_{1,Y} \mu_{2,Y}^n \over  
                    \mu_{1,\rm MR} \mu_{2,\rm MR}^n },
\end{eqnarray}
where $n = 1,2$. Note that $g_{01}$ is negative, so that $K_{\rm FT}$ and 
$K_{\rm LBB}$ are positive as expected.

\end{document}

%% file: TXSsymb.tex
\def\frac#1#2{{#1\over#2}}

\def\half{\ifinner {\scriptstyle {1 \over 2}}%
          \else {\textstyle {1 \over 2}}\fi}

\def\simge{
    \mathrel{\rlap{\raise 0.511ex
        \hbox{$>$}}{\lower 0.511ex \hbox{$\sim$}}}}

\def\simle{
    \mathrel{\rlap{\raise 0.511ex
        \hbox{$<$}}{\lower 0.511ex \hbox{$\sim$}}}}

\def\therefore{
   \setbox0=\hbox{$.\kern.2em.$}\dimen0=\wd0    %
   \mathrel{\rlap{\raise.25ex\hbox to\dimen0{\hfil$\cdotp$\hfil}}%
   \copy0}}

\def\|{\ifmmode\Vert\else \char`\|\fi}

\def\tr{\mathop{\rm tr}\nolimits}       

\def\sc#1{\setbox0=\hbox{$#1$}           
   \dimen0=\wd0                                 
   \setbox1=\hbox{/} \dimen1=\wd1               
   \ifdim\dimen0>\dimen1                        
      \rlap{\hbox to \dimen0{\hfil/\hfil}}      
      #1                                        
   \else                                        
      \rlap{\hbox to \dimen1{\hfil$#1$\hfil}}   
      /                                         
   \fi}                                         %

\def\subrightarrow#1{
  \setbox0=\hbox{
    $\displaystyle\mathop{}
    \limits_{#1}$}
  \dimen0=\wd0
  \advance \dimen0 by .5em
  \mathrel{
    \mathop{\hbox to \dimen0{\rightarrowfill}}
       \limits_{#1}}}                           

\def\vbig#1#2{{\vbigd@men=#2\divide\vbigd@men by 2%
   \hbox{$\left#1\vbox to \vbigd@men{}\right.\n@space$}}}

